\documentclass[12pt]{article}
 \pdfoutput=1
 \usepackage{graphicx}
  \usepackage{hyperref}
  \usepackage{epsfig}
  \usepackage{amsmath}
  \usepackage{amssymb}
  \usepackage{color}
  \usepackage{mciteplus}
  \newlength{\abstractwidth}
  \newcommand{\be}{\begin{equation}}
  \newcommand{\ee}{\end{equation}}
  
  \renewcommand{\title}[1]{\vbox{\center\bf{\Large{#1}}}\vspace{5mm}}
  \renewcommand{\author}[1]{\vbox{\center#1}\vspace{5mm}}
  \newcommand{\address}[1]{\vbox{\center\em#1}}
  \newcommand{\email}[1]{\vbox{\center\tt#1}\vspace{5mm}}
  \definecolor{darkgreen}{rgb}{0,.5,0}

\begin{document}

\begin{titlepage}
\rightline{MIT-CTP 4404}
\rightline{SU-ITP-12/31}
\begin{center}
\hfill \\
\hfill \\
\vskip 1cm

\title{On memory in exponentially expanding spaces}

\author{Daniel A. Roberts${\,}^{a}$ and Douglas Stanford${\,}^{b}$}

\address{$^{a}$ Center for Theoretical Physics {\it and} \\  Department of Physics, Massachusetts Institute of Technology \\
Cambridge, MA 02139, USA \\ $^{b}$ Stanford Institute for Theoretical Physics {\it and} \\ Department of Physics, Stanford University,  Stanford, CA 94305, USA}

\email{$^a$drob@mit.edu,
$^b$salguod@stanford.edu}

\end{center}
  
  \begin{abstract}
  We examine the degree to which fluctuating dynamics on exponentially expanding spaces remember initial conditions. In de Sitter space, the global late-time configuration of a free scalar field always contains information about early fluctuations. By contrast, fluctuations near the boundary of Euclidean Anti-de Sitter may or may not remember conditions in the center, with a transition at $\Delta=d/2$. We connect these results to literature about statistical mechanics on trees and make contact with the observation by Anninos and Denef that the configuration space of a massless dS field exhibits ultrametricity. We extend their analysis to massive fields, finding that preference for isosceles triangles persists as long as $\Delta_- < d/4$.
  \end{abstract}

  \end{titlepage}

\tableofcontents

\baselineskip=17.63pt

\section{Introduction}
The exponential expansion of de Sitter space tends to wash away information about initial conditions \cite{Gibbons:1977mu,*Hawking:1981fz}. This cosmic no-hair principle, which has both classical \cite{Wald:1983ky,*Starobinsky:1982mr} and quantum mechanical \cite{Marolf:2010nz,*Hollands:2010pr} versions, gives inflation \cite{Guth:1980zm,*Linde:1981mu,*Starobinsky:1980te} predictive power, and may do the same for eternal inflation \cite{reviewofmeasure,*Freivogel:2011eg,*Salem:2011qz}.  Cosmic no-hair can be paraphrased as follows: expectation values of quantities defined in a fixed number of regions of fixed proper size forget the initial conditions at sufficiently late time. The ``fixed'' qualifiers are important. Local quantities forget initial conditions, but global quantities, such as integrals of fields over the entire spatial slice, may not. 

A closely related question has been studied thoroughly in the context of statistical mechanics on trees. The prototypical example, reviewed in \cite{broadcasting}, is the Ising model on an infinite tree with free boundary conditions. This system can be written via transfer matrix as a Markov problem, and an analog of cosmic no-hair follows from Markov convergence. The more interesting question, known in the literature as the ``reconstruction'' or ``broadcast'' problem, is whether the probability distribution for a global quantity, such as the total magnetization of the spins on the leaves of the tree, depends on the value of a spin at the root. As it turns out, there is a phase transition. Below a critical $T_c$, the majority vote of spins at infinity tends to coincide with the root \cite{mooresnell,kestenstigum}, while, above $T_c$, the joint distribution for root and boundary spins is exactly a product: all correlation is washed away \cite{onthepurity}. We will describe a system as having ``memory'' if significant correlation between the root and simple global quantities persists as we approach the boundary in the tree.

In Section~\ref{Memory}, we will exploit the geometric similarity of the regular tree, de Sitter space, and Euclidean anti-de Sitter space to define memory and locate the analog of this $T_c$ transition for field theory on dS and EAdS. We will begin in Section~\ref{subsect:isingontree} by reviewing in detail the transition on trees. There, memory depends upon the existence of correlations that decay no faster than the square root of the rate at which the volume of the tree increases. This geometrical criterion is easy to translate into (anti) de Sitter space, where it corresponds to the condition $\Delta \le d/2$, where $\Delta$ is the dimension that characterizes the falloff of correlation functions. 

For free field theory in de Sitter, this criterion is always satisfied, so we expect such fields to always have memory. We confirm this by explicit calculation in Section~\ref{dS}. In fact, the existence of memory for free fields in dS can be understood as a consequence of mode-by-mode unitarity. It is interesting to note that unitarity is made compatible with the $\Delta \le d/2$ criterion by a branch cut in the behavior of $\Delta$ as a function of mass. By contrast, interacting fields in de Sitter can have $\Delta > d/2$, and we expect that simple global quantities forget perturbations to the initial conditions.

In EAdS, free fields can have dimensions greater than or less than $d/2$, depending on the choice of standard ($\Delta > d/2$) or alternate ($\Delta <d/2$) boundary conditions \cite{Breitenlohner:1982jf,*Klebanov:1999tb}, so the criterion above suggests a transition. We confirm this in Section~\ref{EAdSsect}. The fact that this can go either way means that the memory phenomenon should have an interpretation purely in CFT terms. Indeed, one way to make the analogy is to consider a Euclidean CFT perturbed by a relevant operator at infinity. The existence of memory becomes the question of whether the statistics of functions of an order one fraction of all UV degrees of freedom are sensitive to the infrared perturbation.  While somewhat orthogonal to the main direction of this paper, this CFT interpretation is explored in Appendix~\ref{memory-CFT}.

In Section~\ref{Ultrametricity}, we will study the implications of memory for the space of field configurations. To make the point vivid, one can consider the dynamics of nonperturbative bubble nucleation in de Sitter space. This can be characterized by a collection of highly non-Gaussian fields that keep track of the local vacuum index \cite{symmetree}. In the case where the nucleation probabilities are exponentially small, the dimension $\Delta$ of such fields is order $e^{-S_{CDL}}$, and the memory is exceptionally clear: a late-time configuration will contain enough information to very accurately reconstruct the entire history up to that point.


Mathematically, the existence of memory is related to the presence of multiple extreme components in the Gibbs measure for field configurations.\footnote{We will use the term ``extreme state'' instead of ``pure state.'' And we mean extreme states in the tree/bulk, not in the CFT.} A closely related point was recently made by Anninos and Denef \cite{cc}, who analyzed the Hartle Hawking state for a massless field in dS and demonstrated the existence of multiple extreme states. They went on to compute probability distributions for the distances between three field configurations, independently drawn from the Hartle-Hawking ensemble, and found a non-Gaussian, somewhat ultrametric distribution of distances. In Section~\ref{AD} we will extend their analysis to positive mass fields in de Sitter, finding that the Anninos-Denef ultrametricity extends to massive fields as long as $\Delta_- < d/4$, but that the overlap distributions become Gaussian beyond this point.

\section{Global memory of intial conditions}
\label{Memory}
\subsection{Ising model on a tree}\label{subsect:isingontree}
\begin{figure}[ht]
\begin{center}
\includegraphics[scale=0.8]{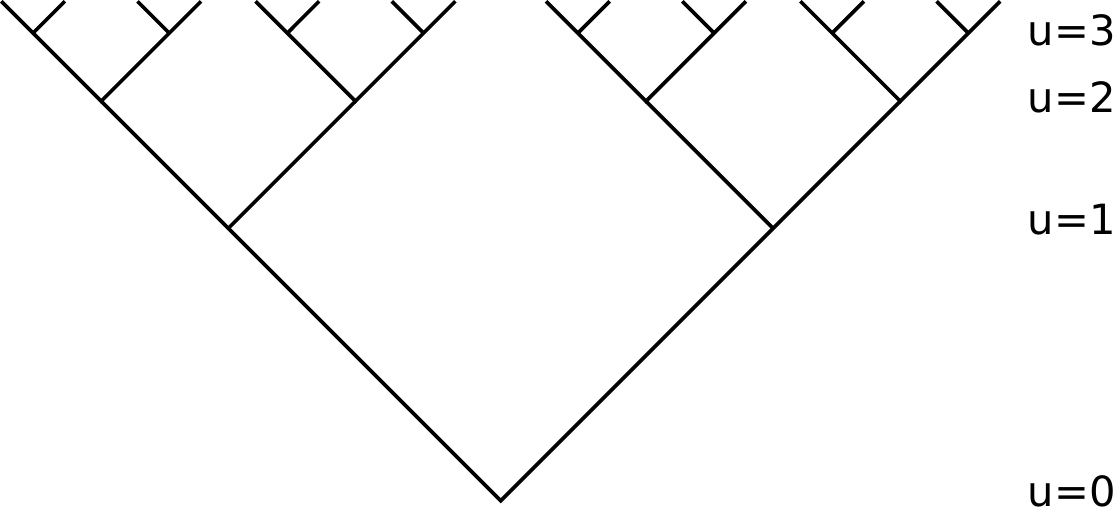}
\end{center}
\caption{The time $u$ in a $p=2$ tree.}
\label{picture-of-tree}
\end{figure}

Like many concepts in statistical mechanics, memory can be illustrated most simply with the Ising model. Specifically, consider the Ising model on a (rooted) regular tree of degree $(p+1)$. This system is defined by associating a classical spin variable $s_i = \pm 1$ to each site, and taking the Boltzmann weighting at temperature $T$ with the usual pairwise Hamiltonian, 
\be
\label{hamiltonian}
H = J\sum_{\text{links }ij} s_is_j.
\ee
Write $\mathbf s^{(u)}$ for the collection of spins at generation $u$ (see Fig.~\ref{picture-of-tree}) in the tree, and consider the mutual information
\be
\label{mutual}
I(\mathbf{ s^{(u)}};\mathbf{ s^{(0)}}) = S(\mathbf{ s^{(u)}}) + S(\mathbf{ s^{(0)}}) - S(\mathbf{ s^{(u)}},\mathbf{ s^{(0)}})
\ee
between the spins at generation $u$ and at the root. If this quantity approaches zero as $u$ tends to infinity, then conditioning on the value of the root spin doesn't change the distribution for the spins at $u\rightarrow \infty$, and we will say that there is no memory. On the other hand, if the mutual information is bounded from below by a positive constant as $u$ tends to infinity, then there is significant correlation between the spin at the root and the collection of spins at $u \rightarrow \infty$, and we will say that there is memory.
A priori, it might have been the case that the Ising model on a tree either always has memory, or never does. Here, we will give a heuristic argument that a transition happens at a finite value of the temperature, leaving the proof to the literature \cite{kestenstigum,mooresnell,onthepurity}.

\begin{figure}[ht]
\begin{center}
\includegraphics[scale=0.35]{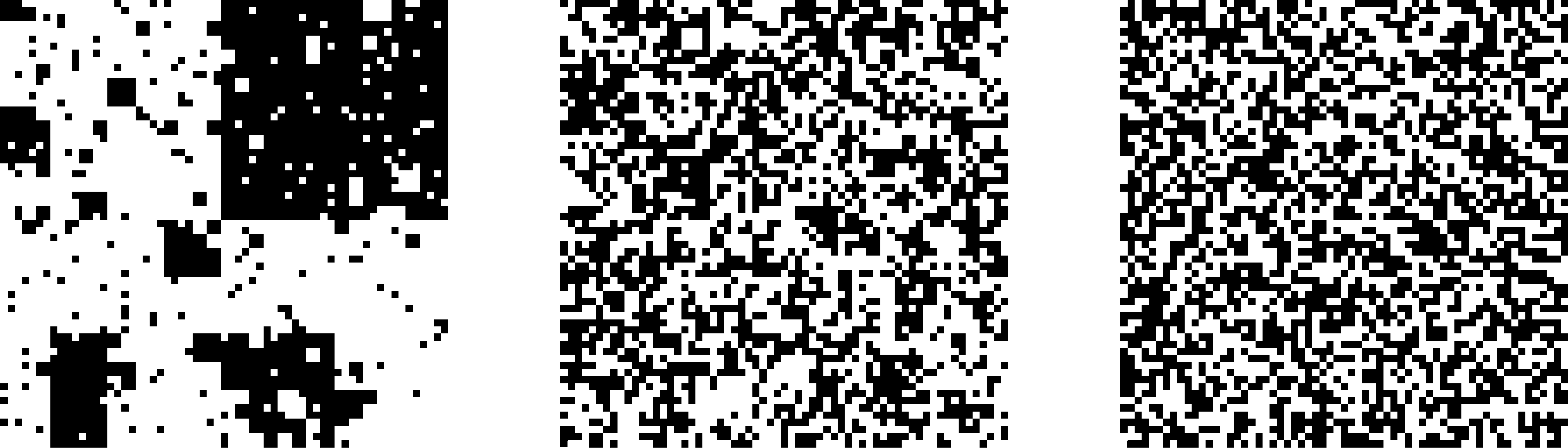}
\end{center}
\caption{Sample configurations of the spins in the $p=4$ tree after six generations, given a white initial condition at generation zero. The left panel is firmly on the ``memory'' side of the transition, with $\lambda^2p = 3.24$ or ``flip probability'' $\gamma=0.05$. The middle panel is at the transition $\lambda^2p = 1$ or $\gamma=0.25$, and the right panel is forgetful, at $\lambda^2p = 0.16$ or  $\gamma=0.4$.}
\label{memory-transition}
\end{figure}

Spin-spin correlation functions in the tree are given by
\begin{align}
\label{corr-ising}
\langle s(x)s(y)\rangle = \lambda^{d(x,y)}, \\
\lambda = \tanh J/T, \notag
\end{align}
where $d(x,y)$ is the number of links separating the vertices $x$ and $y$. In particular, this means that if we impose an initial condition that the root has spin up, the effect on a given spin at generation $u$ will be a transient that decays as $\lambda^u$. For positive temperature, it follows that there is no {\it local} memory of the initial condition. However, let us consider a global quantity: the total magnetization at generation $u$
\be
\label{mag}
M_u = \sum_{i = 1}^{p^u}s^{(u)}_i.
\ee
Here, we are denoting the $i$-th spin at generation $u$ by $s^{(u)}_i$. Note that $M_0 = s_1^{(0)}$, which is the spin of the root. Using Eq.~\eqref{corr-ising}, it is easy to see that
\be
\label{fade}
\langle M_u M_0\rangle =(p\lambda)^u,
\ee
so that the correlation between the total magnetization at generation $u$ and the spin at the root actually grows as a function of $u$, as long as $\lambda p>1$. One might be tempted to conclude that this condition is sufficient for the existence of memory, but we need to be more careful. A large covariance between $M_u$ and $M_0$ does not necessarily mean a large mutual information, since the variance of $M_u$ grows with $u$. However, for this two-state Ising model, one can show that the mutual information is bounded from below by a ``normalized'' correlation function \cite{broadcasting} and from above by a sum of squares of correlations:
\be
\frac{\langle M_u M_0\rangle^2}{\langle M_u^2\rangle} \le I(M_u,M_0) \le \sum_{i=1}^{p^u}\langle s^{(u)}_is^{(0)}_1\rangle^2.
\ee
Using Eq.~\eqref{corr-ising}, we evaluave the upper bound as $\lambda^{2u}p^u$. To evaluate the denominator in the lower bound, we sum over pairs of spins at generation $u$ and apply Eq.~\eqref{corr-ising}. The result is\footnote{Note added in proof: after submission of this version to JHEP, Ref.~\cite{Magan:2013msa} was released, in which the following calculation is also performed.}
\begin{align}
\langle M_u^2\rangle &= p^u\left(1 + \sum_{n=1}^u\lambda^{2n}p^{n-1}(p-1)\right) \notag \\
&=p^u\frac{1-\lambda^2 - \lambda^2(p-1)(\lambda^2p)^{u}}{1-\lambda^2p}.
\end{align}
Combining this with Eq.~\eqref{fade}, we find
\be
\lim_{u\rightarrow\infty}\frac{\langle M_u M_0\rangle^2}{\langle M_u^2\rangle} = \left\{
     \begin{array}{lr}
       0 &  \lambda^2p \le 1 \\
       \frac{\lambda^2p-1}{\lambda^2(p-1)} &  \lambda^2p>1.
     \end{array}
   \right.
\ee
We therefore have a critical value $\lambda = p^{-1/2}$. If $\lambda$ is larger than $p^{-1/2}$, the mutual information $I(M_u,M_0)$ is bounded below by a positive quantity as $u\rightarrow \infty$, and there is memory. If $\lambda$ is less than or equal to $p^{-1/2}$ then the upper bound forces $I(M_u,M_0)$ to zero; there is no memory.  A visualization of this memory transition for the boundary of a $p=4$ tree is show in Fig.~\ref{memory-transition}.

To be more precise, the above argument establishes that the majority-vote variable $M_u$ forgets the initial condition at $u = 0$ if $\lambda < p^{-1/2}$. One might wonder whether other variables built from the spins at infinity might have mutual information with the root, even for $\lambda < p^{-1/2}$. For the specific case of the Ising model, this turns out not to be the case \cite{onthepurity,broadcasting}. However, for more general systems, we can formulate an analogous criterion, defining $\lambda$ by identifying $\lambda^{d(x,y)}$ as the slowest-decaying correlator, and it is known that $\lambda > p^{-1/2}$ is sufficient but not necessary for memory. In the remainder of the paper, we will focus on ``majority vote'' memory---i.e. mutual information between a simple global census at $u\rightarrow\infty$ and the initial condition, for which the condition $\lambda > p^{-1/2}$ is necessary and sufficient \cite{mossel2001reconstruction}.

It is sometimes useful (and very natural) to rewrite the statistical mechanics problem defined by Eq.~\eqref{hamiltonian} as a branching Markov process. From this point of view, the probability distribution for a spin at generation $(u+1)$ depends on the spins at earlier generations only through the parent at generation $u$. Combined with the free boundary conditions at infinity, this means that if parent spin is up, the child is more likely to be up than down by a ratio of Boltzmann factors $e^{2J/T}$. More generally, we can write the probability distribution for the spin of the child as a two component vector, obtained from that of the parent by a Markov matrix
\be
G = \left( \begin{array}{cc}
1-\gamma & \gamma \\
\gamma & 1-\gamma  \end{array} \right),
\ee
where the ``flip probability'' $\gamma$ is given in terms of the temperature by the equation
\be
\frac{P(\text{child }=\text{ parent})}{P(\text{child }\neq\text{ parent})} =  e^{2J/T} = \frac{1-\gamma}{\gamma}.
\ee
To obtain the full probability distribution for all spins at generation $u+1$, one takes the distribution for the parent spins at generation $u$, assigns each parent two children, and applies the matrix $G$ independently for each child. 

In particular, the evolution of the probability distribution along a given path through the tree is an ordinary Markov process. If we start out the root with spin up, the probability distribution for a given child at generation $u$ will be evenly split between up and down, up to an exponentially decaying transient, proportional to $\lambda^u$, where $\lambda = 1-2\gamma$ is the second eigenvalue of the Markov matrix $G$. This is simply a restatement of what we already know: the correlation function of the root with a single spin at level $u$ is $\lambda^u$. 

Also, from this point of view, the lack of memory in a given branch is simply the statement that Markov chains converge. The second eigenvalue $\lambda$ of the Markov matrix  controls the ``burn-in''---i.e. the distance one has to go in a single branch in order to escape the effect of initial conditions. Memory can be thought of as the fact that this ``burn-in'' effect does not necessarily apply globally when the volume of space is expanding exponentially.

In fact, there is a simple mnemonic for this criterion that will be helpful in what follows. At generation $u$, $p^u$ is the number of vertices, which we identify with the volume of space. Meanwhile, $\lambda^u$ is the correlation between a spin at generation $u$ and the root. The condition $\lambda>p^{-1/2}$ for majority-vote memory can thus be phrased as the requirement that the square of the correlation function multiplied by the volume should increase with $u$.



\subsection{de Sitter space}\label{dS}
\begin{figure}[ht]
\begin{center}
\includegraphics[scale=1.6]{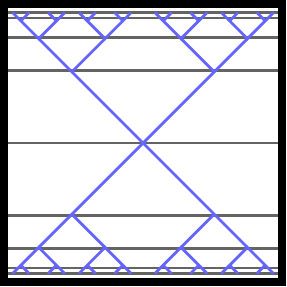}
\end{center}
\caption{The geometry of de Sitter space is analogous to that of a tree. The light grey lines on the Penrose diagram are constant global time $t$.}
\label{dS-figure}
\end{figure}

Geometrically, de Sitter space is very similar to the tree, with $u$ playing the role of proper time $t$ (see Fig.~\ref{dS-figure}).
We will thus define majority vote memory by analogy to the previous section: if the global average of a field $\phi$ at $t\rightarrow \infty$ remains significantly correlated with its value at $t = 0$, then there is memory. Otherwise, there is not.

Does field theory in de Sitter space have a transition analogous to the $\lambda = p^{-1/2}$ transition in the tree? It is straightforward to check for free fields, since the two-point correlation function of a massive scalar is known exactly. In global coordinates, for which the metric is \cite{dsreview,*Bousso:2002fq, *Anninos:2012qw}
\be
ds^2 = -dt^2 + \cosh^2td\Omega_d^2,
\ee
the scalar correlation function is (see, e.g. \cite{Spradlin:2001pw})
\begin{align}
\label{corr}
\langle\phi(t,\theta)\phi(t',0)\rangle=\frac{\Gamma[\Delta_{+}]\Gamma[\Delta_{-}]}{(4\pi)^{(d+1)/2}\Gamma[(d+1)/2]} \ {}_2F_1\left(\Delta_{+},\Delta_{-};\frac{d+1}{2};\frac{1+Z}{2}\right) \\
\Delta_\pm = \frac{1}{2}(d \pm \sqrt{d^2 - 4m^2}) \hspace{20pt} Z=-\sinh t\sinh t' + \cosh t\cosh t'\cos\theta \notag.
\notag
\end{align}
By expanding the hypergeometric function near infinity, we find that the magnitude of the correlator $\langle \phi(t,0)\phi(0,0)\rangle$ decays at late time like $e^{-\text{Re}(\Delta_-) t}$, and we note that Re$(\Delta_-)\le d/2$.
In direct analogy to the total magnetization from the Ising section, Eq.~\eqref{mag}, we define $M_t$ as the zero mode
\be
M_t = \int_{S^d} \phi(\Omega,t)\cosh^dtd\Omega.
\ee
One can check, either using the large $t$ behavior of Eq.~\eqref{corr} or by computing the Green's function for the zero mode directly, that 
\be
\label{shift}
\langle M_tM_0\rangle \sim e^{(d-\Delta_-)t}
\ee
and
\be
\langle M_t^2\rangle \sim e^{2(d-\Delta_-)t}.
\ee
In particular, we find that $\langle M_t M_0\rangle ^2/ \langle M_t^2\rangle$ remains finite in the late-time limit. For Gaussian variables, positivity of this ratio implies mutual information, so we conclude that free fields in dS always have memory, for any value of the mass.\footnote{There is a slight subtlety if $m = d/2$ and logarithms appear in the late-time behavior. It can be checked that the ratio is still order one in the late time limit. If $m > d/2$, then $\Delta_-$ becomes complex, and $\langle M_tM_0\rangle$ oscillates with $t$. However, the ratio of the envelopes of $\langle M_t M_0\rangle ^2$ and $\langle M_t^2\rangle$ is finite in the late-time limit, so significant correlation remains.}

In fact, this is to expected: for free theories, unitarity is a particularly powerful constraint, since the different modes decouple, and the evolution of each mode must separately preserve information. This already implies that, while cosmic no-hair may apply to local quantities, such fields must retain memory in global Fourier modes. 

What happened to the $\lambda = p^{-1/2}$ transition from the tree? Recall the mnemonic from the end of Section~\ref{subsect:isingontree}: the requirement for memory was that the square of the correlation function multiplied by the volume should grow with time. In de Sitter, the spatial volume grows as $\cosh^d t$, while $\langle \phi(t,0)\phi(0,0)\rangle$ decays at late time as $e^{-\Delta_-t}$. The square of the correlator times the volume, then, behaves at late time as $e^{(-2\Delta_- + d)t}$, which is nondecreasing due to the fact that $\text{Re}(\Delta_-) \le d/2$ for any value of the mass. In other words, the unitarity argument is made consistent with the $\lambda > p^{-1/2}$ condition from Section~\ref{subsect:isingontree} by the branch cut in the formula for $\Delta_-$. Either way, we conclude that free fields in de Sitter space always have memory.

On the other hand, interacting theories in de Sitter can have a falloff faster than $\Delta_- = d/2$ \cite{Marolf:2010zp}. It seems likely that, in such theories, perturbations of the initial state are not recorded in the statistics of simple late-time quantities, such as the spatial integrals considered here. Indeed, because interactions mix different modes and ``scramble'' information about initial conditions, unitarity does not imply majority vote memory in the interacting case.

Finally, it is interesting to interpret memory in the context of eternal inflation. The dynamics of non-perturbative bubble nucleation can be represented by a large number of fields with exponentially slow falloff, $\Delta_- \sim e^{-S_{CDL}}$ \cite{symmetree}, or $(1-\lambda) \sim e^{-S_{CDL}}$. For such fields, the correlation between late-time configurations and early fluctuations is extremely strong, allowing a near-perfect reconstruction of the entire previous history from the configuration at a single spatial slice. This suggests that simple observables in dS/CFT \cite{Witten:2001kn,*Strominger:2001pn,Maldacena:2002vr} or FRW/CFT \cite{Freivogel:2006xu} may be sufficient to accurately determine the nucleation history in eternal inflation. We will comment further on the embedding of bulk history in the space of boundary configurations in Section~\ref{Ultrametricity}.

\subsection{Euclidean Anti-de Sitter space} \label{EAdSsect}
Field theory on hyperbolic space (Euclidean AdS) can be viewed as a continuous version of statistical mechanics on a tree graph. Accordingly, the translation of the tree definition of memory is direct: if the mutual information between the field variable at some fixed coordinate in the bulk of EAdS and the collection of field variables near the boundary stays bounded away from zero as we approach the boundary, then we will say that there is memory.  As a warmup, in Appendix~\ref{subsect:gaussiantree} we consider the memory problem using a transfer matrix approach for a massive Gaussian field on a tree, which, qualitatively, has the same behavior as the continuous EAdS case presented below.

The Poincare metric for EAdS is
\begin{equation}
ds^{2}=\frac{dz^{2}+dx^{i}dx^{i}}{z^{2}} \hspace{20pt} i=1,\ldots,d,
\end{equation}
and the action for a massive scalar field $\phi(\mathbf{x},z)$ is
\be
S = \frac{1}{2}\int \frac{d^d\mathbf{x} dz}{z^d}\Big\{z^2 (\partial_z \phi)^2 + z^2(\partial_i\phi)^2 + m^2\phi^2 \Big\}.
\ee
The associated wave equation for the spatially homogeneous mode $\varphi(z)$ has two independent power-law solutions, $z^{\Delta}$ and $z^{d-\Delta}$, where 
\be
\Delta = \frac{1}{2}\left(d+\sqrt{d^{2}+4m^{2}}\right).
\ee
To define the theory, we need to specify a bounday condition at $z = 0$. We will consider two choices, the ``standard'' and ``alternate'' boundary conditions, which set to zero the coefficients of $z^{d-\Delta}$ and $z^{\Delta}$, respectively \cite{Breitenlohner:1982jf,*Klebanov:1999tb}.

To check for memory, we could use the behavior of two-point correlation functions, as in the previous section. For variety, and to emphasize the role of boundary conditions, we will use a different but equivalent approach, working directly with the probability distributions. Specifically, we will impose a boundary condition $\varphi(\ell') = \varphi_{\ell'}$ at a fixed coordinate $z = \ell'$, and then compute the conditional probability for $\varphi(\ell)$ at a coordinate that approaches the boundary $z = \ell\rightarrow 0$. We will diagnose memory by comparing the center and width of this distribution.\footnote{ We could have also used this method to check for memory in dS. Normally, in de Sitter space, one uses the Bunch-Davies ground state wave function, which is computed in Appendix~\ref{deSitterCal}. Instead, here we would change the initial condition by constraining the initial value of the zero mode of the field at some early time and smearing with a Gaussian (otherwise the wave function is pure phase).}

The desired conditional probabilities can be computed by a path integral, with standard or alternate conditions at $z\rightarrow 0$, and Dirichlet conditions $\varphi(\ell') = \varphi_{\ell'}$ and $\varphi(\ell) = \varphi_{\ell}$. In computing these path integrals, we will mimic the approach of \cite{Heemskerk:2010hk}, splitting the computation into a ``UV'' piece $0 < z < \ell$, and a ``visible'' piece, $\ell < z < \ell'$. Up to normalization, the conditional probability is
\begin{equation}\label{uvir}
P(\varphi_{\ell} | \varphi_{\ell'} )=\Psi_{UV}(\varphi_{\ell})\Psi_{VI}(\varphi_{\ell},\varphi_{\ell'}).
\end{equation}
In this expression, $\Psi_{UV}$ depends on the standard/alternate boundary conditions, and we'll handle the two cases separately below. However, $\Psi_{VI}$ is independent of the boundary conditions. It is defined as the path integral over field configurations on $\ell < z < \ell'$, with $\varphi(\ell) = \varphi_{\ell}$ and $\varphi(\ell')=\varphi_{\ell'}$. We can do this Gaussian path integral by evaluating the action on the appropriate classical solution,
\begin{equation}
\Psi_{VI}(\varphi_{\ell},\varphi_{\ell'})\sim\exp\left(-S_{cl}\right).
\end{equation}
Using the equations of motion and integrating by parts, the action reduces to two surface terms. The exact result for the zero mode is
\begin{equation}
S_{cl} = \frac{\Delta}{2}\left(\frac{\varphi_{\ell'}^2}{\ell'^d} -\frac{\varphi_\ell^2}{\ell^d} \right) - \frac{2\Delta-d}{2}\left(\frac{\left(\ell'^\Delta\varphi_\ell - \ell^\Delta\varphi_{\ell'}\right)^2}{\ell^{2\Delta}\ell'^d - \ell^d\ell'^{2\Delta}}\right).
\end{equation}

\subsubsection{Standard quantization}
To compute $\Psi_{UV}$, let us first consider standardard boundary conditions, where $\varphi(z)\to0$ as $z^{\Delta}$. The $UV$ wave function is defined as a path integral over field configurations respecting this condition at the boundary, and matching $\varphi(\ell) = \varphi_\ell$. The result is \cite{Harlow:2011ke}
\begin{equation}
\Psi_{UV}(\varphi_{\ell})\sim\exp\Big\{-\frac{\Delta}{2\ell^{d}}\varphi_{\ell}^{2}\Big\}.
\end{equation}
To find the conditional probability, we take the product of wave functions as in Eq.~\eqref{uvir}, completing the square by adding a term proportional to $\varphi_{\ell'}^2$. The result, to leading order in $\ell/\ell'$, is
\begin{equation}
P(\varphi_{\ell} | \varphi_{\ell'} )\sim\exp\Big\{-\frac{2\Delta-d}{2\ell^{d}}\Big(\varphi_{\ell}-\Big(\frac{\ell}{\ell'}\Big)^{\Delta}\varphi_{\ell'}\Big)^{2}\Big\}.
\end{equation}
To assess whether this distribution has memory or not, we will fix $\ell'$ and let $\ell$ tend to zero, asking whether significant correlation remains between the variables $\varphi_\ell$ and $\varphi_{\ell'}$. It is clear from the distribution that the width for $\varphi_{\ell}$ is proportional to $\sim\ell^{\frac{d}{2}}$, and the shift in the direction of $\varphi_{\ell'}$ scales as $\sim\ell^{\Delta}$. Since $\Delta$ is always greater than or equal to $d/2$, the width becomes large compared to the shift, so the variables lose correlation as $\ell$ tends to zero: there is no memory. 

\subsubsection{Alternate quantization}
Next, we consider the alternate boundary conditions, $\varphi(z)\sim z^{d-\Delta}$. This changes the path integral that defines $\Psi_{UV}$, and we find
\begin{equation}
\Psi_{UV}\sim\exp\Big\{-\frac{d-\Delta}{2\ell^{d}}\varphi_{\ell}^{2}\Big\}.
\end{equation}
We now take the product with $\Psi_{UV}$, and complete the square, freely adding a term proportional to $\propto\varphi_{\ell'}^{2}$. The result, to leading order in $\ell/\ell'$, is
\begin{equation}
P(\varphi_{\ell} | \varphi_{\ell'} )\sim\exp\Big\{-\frac{2\Delta-d}{2\ell^{d}}\Big(\frac{\ell}{\ell'}\Big)^{2\Delta-d}\Big(\varphi_{\ell}-\Big(\frac{\ell}{\ell'}\Big)^{d-\Delta}\varphi_{\ell'}\Big)^{2}\Big\}.\label{eq:alt-wave-function-square-completed}
\end{equation}
This time, the shift towards $\varphi_{\ell'}$ and the width compete: both are order $\sim\ell^{d-\Delta}$. It follows that the variables maintain an order one correlation, even in the limit $\ell\rightarrow 0$. In alternate quantization, we conclude that EAdS has memory.\footnote{One might wonder what happens at the point where $\Delta = d/2$. There, one can check that both ``standard'' and ``alternate'' conditions are forgetful, since the width is proportional to $\ell^{d/2}$, and the shift is proportional to $\ell^{d/2}/\log\ell$.}

 Finally, since we defined memory in terms of fluctuation statistics near the boundary of (EA)dS, we should be able to translate the forgetful transition into CFT terms. This is slightly outside the main line of inquiry of the paper, but is presented in Appendix~\ref{memory-CFT}.

\subsubsection{Mutual information}\label{mutualinfo}
So far, we have treated only the spatial zero mode and been somewhat binary in the distinction between memory and forgetfulness. In this section, we will be a little more quantitative. We'll consider a general Fourier component $\varphi_{\mathbf{k}}(z)$, and compute the mutual information $I(\varphi_{\mathbf{k}}(\ell);\varphi_{\mathbf{k}}(\ell'))$ between the mode at some fixed radius $\ell'$ in the bulk of EAdS, and the same mode, evaluated at a radius $\ell$ that approaches the boundary.

The mutual information of two random variables is defined as the sum of the differential entropies of the marginal distributions, minus the entropy of the joint distribution:
\begin{equation}
I(X;Y)=S(X) + S(Y) - S(X,Y).
\end{equation}
Here, the differential entropy $S$ is defined as the integral of $-p(x) \log p(x)$, and can be positive or negative. To compute the relevant entropies, we need the marginal distributions for $\varphi_{\mathbf{k}}(\ell)$ and $\varphi_{\mathbf{k}}(\ell')$, as well as the joint distribution for both. To compute the marginal distribution, we divide the bulk path integral up into IR and UV pieces, and take
\be
P(\varphi_{\mathbf{k}}(\ell)) = \mathcal{N} \Psi_{UV}(\varphi_{\mathbf{k}}(\ell))\Psi_{IR}(\varphi_{\mathbf{k}}(\ell)).
\ee
Here, $\mathcal{N}$ is a normalization constant, and the $IR$ wave function is a path integral over field configurations that are smooth in the interior of the space, and match onto $\varphi_{\mathbf{k}}(\ell)$ at radius $\ell$. As always, we evaluate the various path integrals by evaluating the action on the relevant classical solutions, which are linear combinations of the Bessel functions $I_{\pm\nu}(kz)$, where
\be
\nu = \Delta - \frac{d}{2} = \frac{1}{2}\sqrt{d^2 +4m^2}.
\ee
A straightforward but somewhat tedious evaluation gives the marginal distribution as
\be
P(\varphi_{\mathbf{k}}(\ell))=\mathcal{N}\exp\left\{ -\frac{1}{2\ell^{d}}\frac{\varphi_{\mathbf{k}}(\ell)^{2}}{I_{\pm\nu}(k\ell)K_{\nu}(k\ell)}\right\}  \hspace{20pt}
\ee
where ``$+$'' correponds to standard boundary conditions, and ``$-$'' corresponds to alternate.
\begin{figure}[ht]
\begin{center}
\includegraphics[scale=0.3]{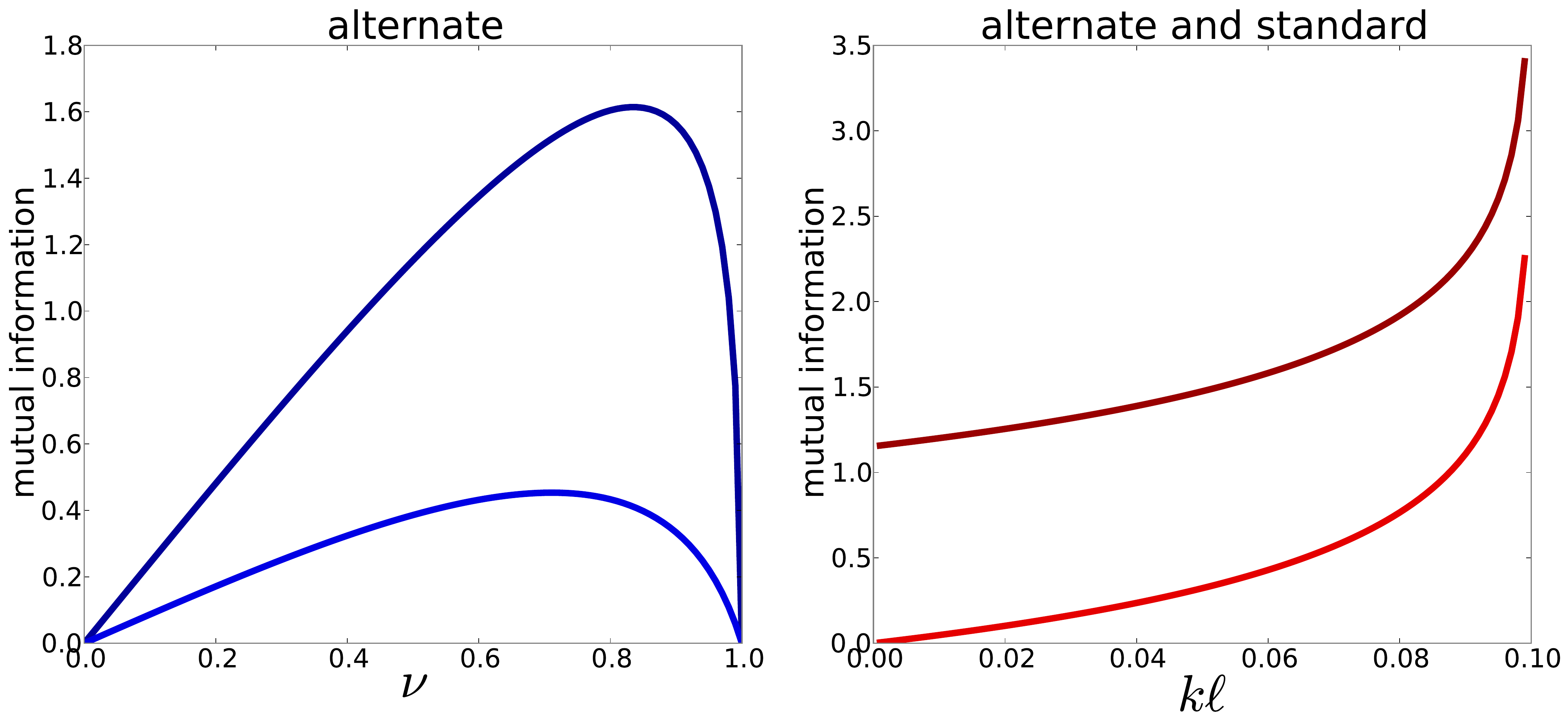}
\end{center}
\caption{(Left) the mutual information $I\big(\varphi_{\mathbf{k}}(\ell);\varphi_{\mathbf{k}}(\ell')\big)$ for alternate quantization, considered as a function of $\nu$. In this plot, $\ell$ has been taken to zero, while $k\ell' = 0.1$ in the top line, and $0.5$ in the bottom line. (Right) the mutual information for both boundary conditions, with $\nu$ fixed at $0.5$ and $k\ell'$ fixed at $0.1$, plotted as a function of $k\ell$. The upper line is alternate, and the lower is standard.}
\label{mutual-fig}
\end{figure}

The expression for the joint distribution is slightly more complicated. In addition to the $UV$ and $IR$ wave functions, one has to compute a $VI$ wave function that implements the path integral over $\ell < z < \ell'$, and then take the product of all three,
\be
P\big(\varphi_{\mathbf{k}}(\ell),\varphi_{\mathbf{k}}(\ell')\big)=\mathcal{N}\Psi_{UV}\big(\varphi_{\mathbf{k}}(\ell)\big)\Psi_{VI}\big(\varphi_{\mathbf{k}}(\ell),\varphi_{\mathbf{k}}(\ell')\big)\Psi_{IR}\big(\varphi_{\mathbf{k}}(\ell')\big).
\ee
This joint distribution is Gaussian, but the covariance matrix is an unpleasant combination of Bessel functions. The mutual information, however, simplifies rather nicely. We find
\be
I\big(\varphi_{\mathbf{k}}(\ell);\varphi_{\mathbf{k}}(\ell')\big)=-\frac{1}{2}\log\left(1-\frac{I_{\pm\nu}\left(k\ell\right)K_{\nu}\left(k\ell'\right)}{I_{\pm\nu}\left(k\ell'\right)K_{\nu}\left(k\ell\right)}\right),
\ee
where, again, we have ``+'' for standard and ``$-$'' for alternate. In the limit of small $\ell$, the mutual information for regular boundary conditions tends to zero as $(k\ell)^{2\nu}$, as shown in the right panel of Fig.~\ref{mutual-fig}. On the other hand, with alternate boundary conditions, we find a nonzero value of $I$ at $k\ell = 0$, plotted as a function of $\nu$ in the left panel of Fig.~\ref{mutual-fig}. The mutual information vanishes at both ends of the alternate quantization window $0 \le \nu \le 1$.

\section{Configuration-space ultrametricity}
\label{Ultrametricity}
In this section, we will review the connection between memory and the existence of multiple extreme (pure) components in the Gibbs state. We will discuss the branching structure of the space of pure states for the Ising model. We will then switch to de Sitter, where Anninos and Denef argued that, similarly, the Bunch-Davies vacuum splits into a tree-like space of extreme states \cite{cc}. We'll generalize their analysis to positive mass.

\subsection{Ising model on a tree}
Mathematically, memory in the Ising model is connected to the existence of nontrivial variables ``at infinity'' in the tree. As an example, we can consider the appropriately normalized total magnetization at generation $u$. For $\lambda > p^{-1/2}$, the limit theorems of \cite{kestenstigum} (see \cite{mooresnell} or \cite{mosselperes} for explanation) establish that the variables
\be
\frac{M_u}{(p\lambda)^u}
\ee
converge to a random variable $M_\infty$, which is correlated with the spin at the root of the tree. Let us contrast this with the case in the no-memory phase $\lambda \le p^{-1/2}$. There, the marginal distributions for the variables
\be
\frac{M_u}{p^{u/2}}
\ee
converge to a Gaussian with mean zero, but the variables themselves do not converge: in a given realization of the spin system, the variable $M_u$ would continue switching sign as a function of $u$.

In probability theory, the concept of variables at infinity is formalized as the tail field, which is the set of observables in an infinite system that don't depend on the variables at any finite number of sites. A system is descrbed as having a trivial tail if every tail event has probability either zero or one. The existence, in the memory phase, of correlation between the variable $M_\infty$ and the root implies that the distribution for $M_\infty$ has finite width, so the tail field is nontrivial. This is equivalent (see \cite{georgii}, Theorem 7.7) to the statement that the free boundary Gibbs state is not an extreme point in the convex space of Gibbs states.\footnote{Extreme (pure) states, and their connection with spin glasses, were recently reviewed in \cite{denef}.} 

We would like to characterize the space of the extreme Gibbs states. For the Ising model, we can get a fairly complete picture of the space of these as follows. First, we divide up the ensemble according to whether the variable $M_\infty$ is positive or negative, in other words, we divide it up according to whether the reconstruction of the initial spin is up or down. This cleanly divides the set of extreme components that makes up Gibbs state into two components. We can repeat this procedure for tail variables corresponding to the total magnetization associated to the leaves of the subtrees emanating from the $p$ children of the root. Focus on the $p = 2$ case for simplicity. The root has two children at level $u = 1$, which we'll label $(1,1)$ and $(1,2)$. Let $M^{(1,1)}_\infty$ and $M^{(1,2)}_\infty$ denote the rescaled magnetization at infinity for the subtrees associated to these vertices. If $M_\infty$ is positive, then there are three possibilities: both $M^{(1,1)}_\infty$ and $M^{(1,2)}_\infty$ might be positive, or they could have opposite sign. Similarly, if $M_\infty$ is negative, then both might be negative or they might have opposite signs. Proceeding in this way, we can further divide the Gibbs state according to the $2^{p^u}$ possible signs of the $p^u$ variables $M^{(u,i)}_\infty$, $i = 1,...,p^u$, where $M^{(u,i)}_\infty$ is the total magnetization at infinity in the subtrees growing out of the $i$-th vertex at generation $u$.

As a function of $u$, this decomposition defines a branching, RG-type evolution of the space of pure states. Because of the existence of memory, the sign of the variable $M^{(u,i)}_\infty$ is correlated with the $i$-th spin at generation $u$ in the tree, so this evolution is related to the evolution of the configurations in the actual system. In the limit of small $\gamma$, this relation becomes exact, but at finite $\gamma$ it is approximate, since the reconstruction of the spin $(u,i)$ from $M^{(u,i)}_\infty$ can be wrong.

\subsection{de Sitter space}\label{AD}
Anninos and Denef recently suggested a very similar extreme state decomposition for the Bunch-Davies vacuum associated to a massless field in de Sitter \cite{cc}. Related overlap distributions were also computed in \cite{Benna:2011as}. Part of the motivation for invoking extreme states was the failure of the massless field to cluster; the two point correlator is logarithmically divergent at large distance. By contrast, the two point function of free massive fields clusters, so one might guess that the analysis of \cite{cc} is related to a pathology of the massless scalar. This is not the case. In the remainder of this section, we will compute overlap distributions for positive mass fields in de Sitter. We'll see that the ultrametric tendency of the massless field is shared by massive fields as long as $\Delta_- < d/4$.

\subsubsection{Setup}
As in section , we will work with de Sitter space in flat slicing, but we'll make the IR cutoff explicit by compactifying space on a torus of comoving size $L$, $x^i \sim x^i + L$. The probability distribution on spatial field configurations is given by the norm-squared of the wave function, $|\Psi|^2$. Following Anninos and Denef, we will define a distance on the space of these configurations,
\be
d(1,2) = \frac{1}{L^d}\int d^dx \left(\hat{\phi}_1(\mathbf{x}) -\hat{\phi}_2(\mathbf{x})\right)^2,
\ee
where $\hat{\phi}$ is obtained from $\phi$ by subtracting the zero mode and smearing over a comoving scale corresponding to a large but fixed number of horizons. It will be convenient to define a regulated distance $\delta(1,2)$ by subtracting the mean,
\be
\delta(1,2) = d(1,2) - \langle d(1,2)\rangle.
\ee
Of course, this subtracted distance can be either positive or negative.

We will be interested in the following questions \cite{cc}: what is the probability, in the ensemble of fluctuations determined by $\Psi$, that two independently chosen configurations have a given distance $\delta$? Or that three configurations will have distances $\delta_1,\delta_2,\delta_3$? Rather than computing the probability distribution for $\delta(1,2)$ directly, it is easier to compute exponential moments $\langle e^{-s\delta(1,2)}\rangle$ over field configurations $\phi_1$ and $\phi_2$, as a function of $s$, and recover the distribution for $\delta$ by inverse Laplace transform
\be
P_\eta(\delta) \propto \int_{-i\infty}^{i\infty}e^{s \delta}\langle e^{-s \delta(1,2)}\rangle_\eta ds.
\ee
The expectation value $\langle\cdot\rangle_\eta$ is done with respect to the measure $|\Psi|^2$ at time $\eta$, provided by the Bunch-Davies wave function. This wave function is computed in Appendix~\ref{deSitterCal}, and the result, for superhorizon modes $k\eta \ll 1$, is
\be
\left|\Psi\big(\phi,\eta\big)\right|^2 = \mathcal{N} \ \exp\Big(-2\sum_{\mathbf{k}}\beta(k,\eta)|\varphi_{\mathbf{k}}|^2\Big)
\ee
where $\mathcal{N}$ is a normalization factor, independent of $\phi$, $k = |\mathbf{k}|$ and
\be
\beta(k,\eta) \sim \frac{L^d\ell_{dS}^{d-1}}{\eta^{2\Delta_-}}k^{d-2\Delta_-} \hspace{20pt} \Delta_- = \frac{1}{2}\left(d - \sqrt{d^2 - 4m^2\ell_{dS}^2}\right).
\ee
The proportionality constant in the definition of $\beta$ is an order one number that depends on $m\ell_{dS}$ and $d$. It is given in the appendix but won't be needed here.

The wave function is a product over the different $\mathbf{k}$ modes, and the distance distribution is quadratic in $\phi_1$ and $\phi_2$, so we can compute the expectation value $e^{-s\delta(1,2)}$ by Gaussian integration, mode by mode. The computation is entirely parallel to the massless one detailed in \cite{cc}, and the result is
\be
\label{newdist}
\langle e^{-s\delta(1,2)}\rangle_\eta = \prod_{\mathbf{k}\neq0}{}^{'} \frac{e^{s/\beta(k,\eta)}}{1+s/\beta(k,\eta)},
\ee
where the primed product runs over unordered pairs $(\mathbf{k},\mathbf{-k})$ with $\mathbf{k}\neq0$. Similarly, the probability distribution for the distances between three configurations is given by a three dimensional Laplace transform of 
\begin{align}
&\langle e^{-s_1\delta(1,2) -s_2\delta(1,3) - s_3\delta(2,3)}\rangle_\eta \notag \\
&\hspace{20pt}= \prod_{\mathbf{k}\neq0}{}^{'} \frac{e^{(s_1+s_2+s_3)/\beta(k,\eta)}}{1+(s_1+s_2+s_3)/\beta(k,\eta)+3(s_1s_2+s_1s_3+s_2s_3)/4\beta(k,\eta)^2}.
\end{align}

If the mass is zero, then the Gaussian kernel $\beta$ is independent of conformal time $\eta$ and we can take the late-time limit safely. However, in the massive case, $\Delta_- >0$, and $\beta$ blows up at late time. This means that the exponential moments tend to zero for any fixed $s$, and the distance distribution collapses to a delta function.

This is a reflection of the fact that massive fields in de Sitter space have a power law fade at late time, proportional to $\eta^{\Delta_-}$. We can compensate for this by defining a new, $\eta$-dependent metric on field configurations,\footnote{For $\Delta_- = d/4$, an additional factor of $1/\log(L/\eta)$ is required in the normalization.}
\begin{align}
\label{metric}
d_{\Delta}(1,2) &= \begin{cases}
\left(\frac{L}{\eta}\right)^{2\Delta_-}\frac{c}{L^d}\int d^d\mathbf{x} \left(\hat{\phi}_1(\mathbf{x}) -\hat{\phi}_2(\mathbf{x})\right)^2  & \text{for } \Delta_- < \frac{d}{4} \\
\left(\frac{L}{\eta}\right)^{d/2}\frac{c'}{L^d}\int d^d\mathbf{x} \left(\hat{\phi}_1(\mathbf{x}) -\hat{\phi}_2(\mathbf{x})\right)^2  & \text{for } \Delta_- > \frac{d}{4}
\end{cases} \notag
\\
\delta_{\Delta}(1,2) &= d_\Delta(1,2) - \langle d_\Delta(1,2)\rangle.
\end{align}
The reason for the change in behavior of the normalization at $\Delta_- = d/4$ will be made clear below, where we'll see explicitly that the above definition ensures that the width of the distribution for $\delta_\Delta(1,2)$ has a finite and nonzero late-time limit. In what follows, we'll adjust the constants of proportionality $c,c'$ as a function of mass so that the variance is one.

\subsubsection{Ultra-light fields}
We'll begin by considering the case $\Delta_- < d/4$, corresponding to a very small mass $(m\ell_{dS})^2 < 3d^2/16$. In this mass range, the explicit power of conformal time in the definition of $\delta_\Delta(1,2)$ cancels the time-dependence of $\beta$. Up to a constant multiple in the definition of $\delta_\Delta$, which we fix by measuring distances in units of the variance, we have
\be
\label{product}
\langle e^{-s\delta_\Delta(1,2)}\rangle = \prod_{\mathbf{n}\neq0}{}^{'} \frac{e^{s/n^{d-2\Delta_-}}}{1+s/n^{d-2\Delta_-}},
\ee
where the primed product runs over unordered pairs $(\mathbf{n},\mathbf{-n})$, and $n = |\mathbf{n}|$. As long as $\Delta_-< \frac{d}{4}$, this product converges, and we are able to remove the smearing function that cuts off high momentum modes. A similar formula holds for the triple overlap. 

As far as we know, this infinite product Eq.~\eqref{product} is not known in closed form. One could approximate the product as the exponential of the integral of the logarithm, which can be evaluated in terms of known functions. But, even with this simplification, it doesn't seem possible to perform the Laplace transform to recover $P(\delta)$, even in the saddle point approximation.

However, even without doing the relevant integrals, it is clear that the distributions are continuous in $\Delta_-$ near $0$, so that the overlap distributions for ultra-light fields match smoothly to the massless distributions as $m\ell_{dS} \rightarrow 0$. This establishes that the ultrametricity of \cite{cc} extends to very small mass.

\begin{figure}[ht]
\begin{center}
\includegraphics[scale=0.22]{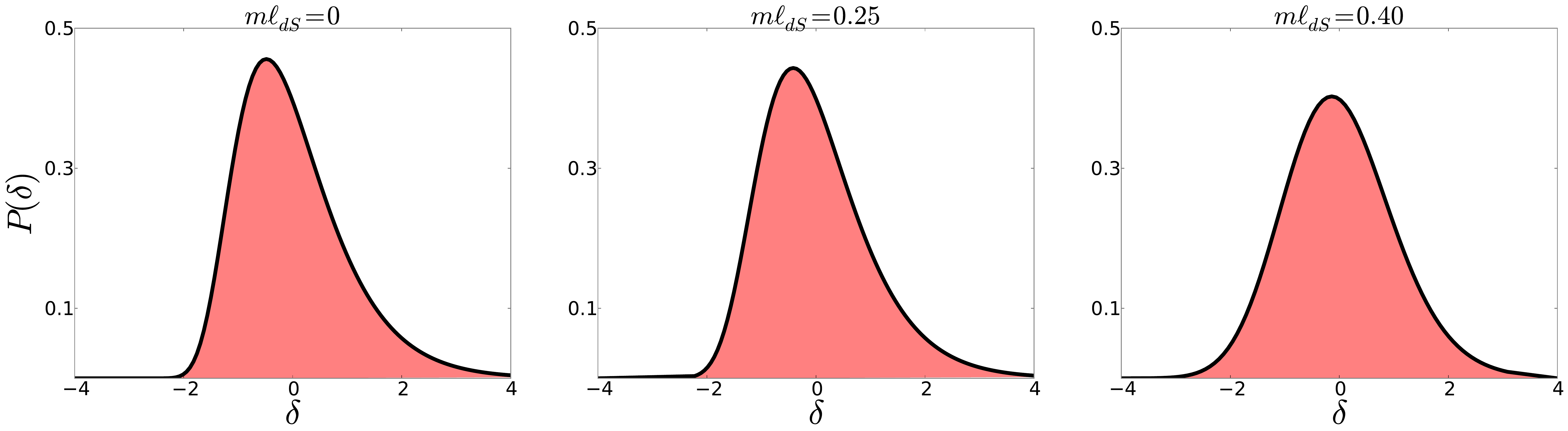}
\end{center}
\caption{Saddle point approximations of dS$_{1+1}$ overlap distributions, measured in units of the variance, for three different values of the mass. $\Delta_- = d/4$ corresponds to $m\ell_{dS} = 0.43$.}
\label{fig1}
\end{figure}

Also, we can get a qualitative picture by studying the distributions numerically. The products are easy to compute, but the oscillatory Laplace transform is inconvenient. Instead, we do the transform approximately by numerically searching for saddle points. As a first example, dS$_{1+1}$ probability distributions for a single distance $\delta_\Delta(1,2)$ are shown in Fig.~\ref{fig1}.
For zero mass, we recover the Gumbel distribution of \cite{cc}. As the mass-squared increases towards the critical value $3d^2/16$, we see the lopsidedness fading as the Gumbel turns into a Gaussian.

\begin{figure}[ht]
\begin{center}
\includegraphics[scale=0.22]{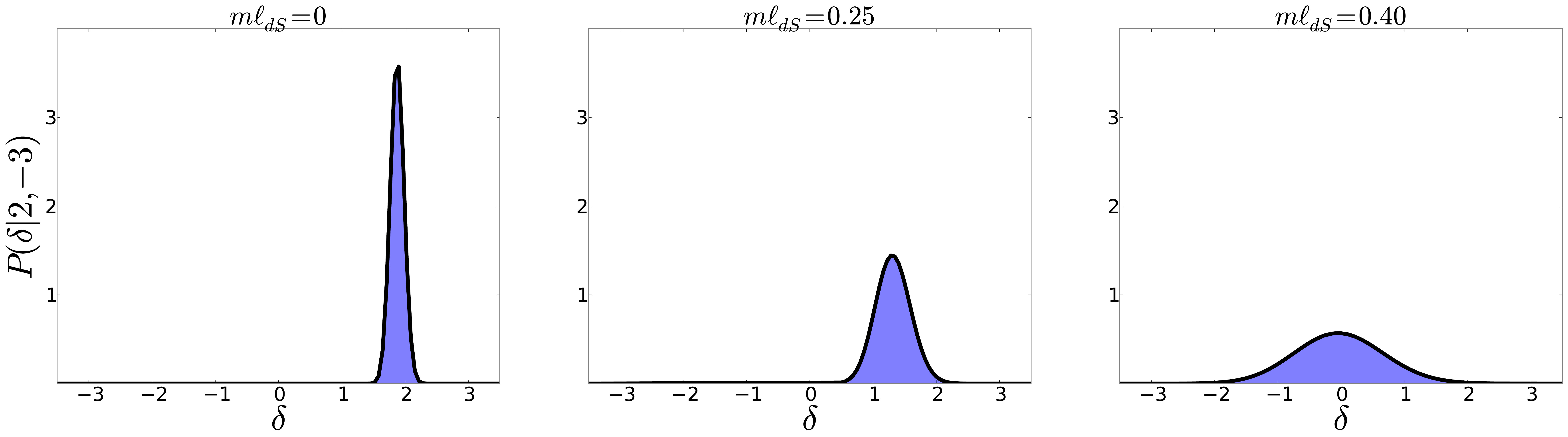}
\end{center}
\caption{Saddle point approximations of dS$_{1+1}$ conditional probability $P(\delta|2,-3)$, measured in units of the variance of the corresponding double overlap distributions, for three different values of the mass. Again, the critical mass is $m\ell_{dS} = 0.43$.}
\label{fig2}
\end{figure}

To check for ultrametricity, we need to evaluate the triple overlap distribution, $P(\delta_1,\delta_2,\delta_3)$. The ultrametricity of \cite{cc} shows itself when one distance is small, by preferring that the other two distances be equal. To check for this behavior, we plot a conditional probability, $P(\delta \ | \ 2,-3)$ in Fig.~\ref{fig2}.
If the distribution were truly ultrametric, this would be a delta function enforcing $\delta=2$. And, indeed, for zero mass, the distribution is rather peaked near $\delta = 2$. As the mass increases from zero, the peak near $\delta=2$ broadens and moves left towards some kind of non-ultrametric compromise between -3, 0, and 2.

It is worth emphasizing that the the conditional probability plotted is {\it very} conditional, in the sense that the absolute probability for having any of the three distances equal to -3 is extremely small. This is apparent in Fig.~\ref{fig1}. To see the sharp ultrametric peaking, we are forced to evaluate the triple overlap distribution in a very rare region of parameter space. If, instead, we were to plot $P(\delta|2,-1)$, we would find little or no evidence of ultrametricity. 

\subsubsection{Heavier fields}
We now turn to non-ultra-light fields, for which $\Delta_- \ge d/4$. For such fields, the scaling factor in Eq.~\eqref{metric} has a different form. The reason is that the infinite product Eq.~\eqref{product} diverges, so we can't naively remove the smearing cutoff on the field modes. Instead, we regulate the product with a comoving cutoff on momentum that is a large but fixed multiple of $1/\eta$. With this prescription, one can check that the definition Eq.~\eqref{metric} ensures a finite late-time limit for the distance distribution. 

In fact, the calculation simplifies rather dramatically, because the product is dominated by the most ultraviolet modes. In the late-time limit, we find 
\be
\langle e^{-s\delta_\Delta(1,2)}\rangle = e^{s^2/2},
\ee
and a similar Gaussian formula for the moments of the triple-overlap distribution. The Laplace transforms are simple, giving $P(\delta)$ as a Gaussian. Normalizing the variance to one, we find the triple-overlap distribution 
\be
P_{\Delta_- \ge d/4}(\delta_1,\delta_2,\delta_3) = \frac{2}{3\sqrt{3\pi^3}}\exp\left(-\frac{5}{9}\left(\delta_1^2+\delta_2^2+\delta_3^2\right) + \frac{2}{9}\left(\delta_1\delta_2+\delta_1\delta_3+\delta_2\delta_3\right)\right).
\ee
The conditional probability for $\delta_1$, given $\delta_2$ and $\delta_3$ is a Gaussian, peaked at $\delta_1 = \frac{\delta_2+\delta_3}{5}$. In particular, if $\delta_2 = -\delta_3$, the conditional probability for $\delta_1$ is symmetric and peaked at zero. This lack of attraction towards the larger of the two other distances provides a sharp criterion for the absence of ultrametricity for $\Delta_- \ge d/4$. Based on the first half of this paper, one might expect the transition to happen at $\Delta_- = d/2$, not $\Delta_- = d/4$. Apparently, memory is necessary for non-Gaussianity of the Anninos-Denef overlap distributions, but not sufficient.\footnote{Numerical exploration suggests that this conclusion is robust to changing the distance metric used to e.g. an $L_1$ instead of an $L_2$ norm.}

\section{Conclusion}
Memory is defined as the existence of correlation between global variables at infinity and local variables at some finite point. A well-studied memory/forgetfulness transition happens in the Ising model on the tree, suggesting a critical value of the dimension $\Delta = d/2$ for fluctuating dynamics in (EA)dS. Indeed, we found:
\begin{itemize}
\item Free fields in dS never fall off faster than $\Delta = d/2$. Despite local cosmic no-hair, global memory always exists. The fact that memory is very accurate for fields with slow falloff may facilitate in answering measure problem questions in dS/CFT or FRW/CFT.
\item Free fields in EAdS with standard quantization have $\Delta \ge d/2$ and forget perturbations deep in the bulk. However, with alternate quantization, the dimension can be less than $d/2$, and global variables at the boundary remain sensitive to such perturbations.
\end{itemize}
We discussed the extreme states implied by the existence of memory in such systems. In de Sitter, we find that the ultrametric structure of these states persists to finite positive mass but disappears at $\Delta_- = d/4$. Memory is necessary but not sufficient for ultrametricity.

\section*{Acknowledgments}
We would like to thank Dionysios Anninos, Frederik Denef, and Ben Freivogel for insights about ultrametricity at positive mass, Max Kleiman-Weiner for discussion of reconstruction, Persi Diaconis and Allan Sly for help with the mathematical literature, and Stephen Shenker for comments on the CFT interpretation. We are grateful to Dionysios Anninos and Frederik Denef for organizing ``Cosmology and Complexity 2012,'' the productive and refreshing workshop at which this work began. D.S. supported by the NSF under the GRF program, by NSF grant 0756174 and by the Stanford Institute for Theoretical Physics. D.A.R. is supported by the DOD through the NDSEG Program, by the Fannie and John Hertz Foundation, and is grateful to the SITP for hospitality during the final stages of this work. This work was supported in part by the U.S. Department of Energy under cooperative research agreement Contract Number DE-FG02-05ER41360.

\appendix

\section{Transfer matrix solution of the Gaussian model on a tree}\label{subsect:gaussiantree}
In this appendix, we discuss the memory transition for a Gaussian field on a tree. We clarify the relation between the falloff of the correlation functions and the eigenvalues of the transfer matrix. A massless Gaussian field on a regular $(p+1)$ tree was previously studied by Zabrodin \cite{zabrodin}. We will focus on the $p=2$ case, but add a mass. To define the model, associate to each vertex $a$ of the tree a real field variable $\phi(a)$, and form the Bolztmann ensemble with temperature $T = 1$ and Hamiltonian (or Euclidean action)
\be
H = \frac{1}{2}\sum_{\text{links}} \left(\delta\phi\right)^2 + \frac{m^2}{2}\sum_{\text{vertices}} \phi^2.
\ee
Just like the Laplacian on hyperbolic space, the discrete tree Laplacian has a gap in the spectrum, so the statistical mechanics of this system makes sense with somewhat negative mass-squared. For the $p = 2$ tree, the gap is $3 - 2\sqrt{2}$, so the action is positive as long as $m^2 > 2\sqrt{2} - 3$. When $m^2$ is negative, we will call it $-\mu^2$.

To get a feel for this system, we can study the equations of motion, obtained by differentiating with respect to $\phi$ at a particular vertex $a$ in the tree. This gives
\be
3\phi(a) - \phi(\text{child}_1) - \phi(\text{child}_2) - \phi(\text{parent}) + m^2 \phi(a) =0.
\ee
If we take a homogeneous ansatz $\phi \sim \lambda^u$, where $u$ is time in the tree, we find two solutions,  
\be
\lambda_\pm = \frac{3+m^2 \pm \sqrt{1+6m^2 + m^4}}{4}.
\ee
For small $m^2$, the solutions are $\lambda_+ = 1+m^2$ and $\lambda_- = (1-m^2)/2$. $\lambda_+$ is greater than one for positive $m^2$, so one of the solutions grows exponentially in the direction of the tree's branching. On the other hand, if $m^2<0$, then both branches are less than one.

It is well known that boundary conditions can be extremely important for statistical mechanics on trees. The reason is that the boundary makes up an order one fraction of the tree, for any cutoff. To be careful, we will treat the boundary vertices separately, modifying the above Hamiltonian to
\be
\label{action}
H = \frac{1}{2}\sum_{\text{links}} \left(\delta\phi\right)^2 + \frac{m^2}{2}\sum_{\text{bulk vertices}} \phi^2 + \frac{m_\partial^2}{2}\sum_{\text{bdry vertices}}\phi^2.
\ee
where, in general, $m_\partial \neq m$. In order to preserve the symmetry of the tree, we would like $m_\partial$ to be independent of the cutoff, in the sense that integrating out the boundary vertices leaves an action for a smaller region, with the same value of $m_\partial$. It is easy to show that this condition allows two solutions as a function of $m^2$. They can be conveniently written in terms of the solutions for $\lambda_\pm$ as 
\be
\label{brdymass}
m_\partial^2 = \frac{1-\lambda_\pm}{\lambda_\pm}.
\ee
For for the special case of $m^2 = 0$, the solutions are $m_\partial^2 = 0$, corresponding to free boundary conditions, and $m_\partial^2 = 1$, corresponding to a condition that tends to suppress fluctuations. More generally, we will see that the ``-'' branch closely parallels the standard boundary conditions in EAdS, while the ``+'' branch resembles alternate boundary conditions.

Much like the Ising model, this system can be recast as a branching Markov random field. The rate matrix $G$ along each link of the tree is a Gaussian kernel
\be
G(\phi,\phi') \sim \exp\Big\{ -\frac{(\phi-\phi')^2}{2} - \frac{\alpha}{2} \phi^2 + \frac{\beta}{2} \phi'^2\Big\}.
\ee
where we'll derive $\alpha$ and $\beta$ below. This kernel assignes the probability distribution for the field value $\phi$ of a child vertex at generation $(u+1)$ in terms of the probability distribution for the parent at generation $u$, via
\be
P_{u+1}(\phi) = \int G(\phi,\phi')P_u(\phi')d\phi'.
\ee
To identify the correct values of $\alpha$ and $\beta$, we require that the infinite product of $G(\phi,\phi')$ along each link in the graph should equal $e^{-H}$. This means that we need $\alpha - 2\beta = m^2$. We also require that the probability stay normalized, which sets $\beta = \alpha/(1+\alpha)$. These two equations determine $\alpha$ and $\beta$ in terms of $m^2$. The equation is quadratic, so we have a choice of two solutions. Again, these can be parameterized in terms of $\lambda_\pm$ as $\alpha = (1-\lambda_\pm)/\lambda_\pm$, and $\beta = 1-\lambda_\pm$. Since the boundary weighting implied by the Markov kernel $G$ is $m_\partial^2 = \alpha$, we see that the upper/lower sign choice here is the same as the corresponding choice for the branch of $m_\partial^2$.

We conclude that the properly normalized Markov kernel is 
\be
G(\phi,\phi') = \frac{1}{\sqrt{2\pi\lambda_\pm}}\exp\Big\{ -\frac{(\phi-\phi')^2}{2} -\frac{1-\lambda_\pm}{2\lambda_\pm} \phi^2 +\frac{1-\lambda_\pm}{2}. \phi'^2\Big\}.
\ee
Our lesson from the previous section was that, to look for memory, we should compute the second eigenvalue of $G$. While $G$ isn't a symmetric matrix, it does satisfy detailed balance, i.e. $G = ZSZ^{-1}$, where $Z$ is diagonal and $S$ is symmetric, so we are guaranteed to have eigenvectors and real eigenvalues. These are 
\begin{align}
&\int G(\phi,\phi')f_n(\phi') d\phi' = \lambda_\pm^n f_n(\phi) \notag \\
&f_n(\phi)  = e^{-a\phi^2}H_n(\sqrt{a}\phi) \hspace{20pt} a = \frac{\lambda_\pm}{2} + \frac{1}{2\lambda_\pm}.
\end{align}
Here, $H_n$ is the (physics convention) Hermite polynomial. One can easily check the above using the representation $H_n(x) = \frac{n!}{2\pi i} \oint \frac{dt}{t^{n+1}}e^{-t^2 + 2xt}$ and using Gaussian integration under the contour integral.

With this solution in hand, there are several points to be made. First, it is clear that the second eigenvalue, for either choice of boundary conditions, is equal to the corresponding falloff $\lambda_\pm$ from the tree equation of motion. Second, if $m^2$ is positive and we pick the ``$+$'' branch, the eigenvalues of $G$ are not bounded. This reflects the familiar fact that alternate quantization doesn't make sense for positive mass-squared. Here, we see it as a breakdown of the normalizability of the Markov matrix. Finally, and most important for our purposes, the ``$+$'' branch always has a second eigenvalue greater than or equal to the critical value $1/\sqrt{2}$,\footnote{Remember, we have specialized to $p = 2$ in this subsection.} while the ``$-$'' branch always has an eigenvalue smaller than or equal to that value. It follows from the theorems in \cite{kestenstigum,mooresnell} that the ``$+$'' version has majority vote memory, and the ``$-$'' version does not.

\section{CFT}\label{memory-CFT}
In this appendix, we recast the memory transition in purely CFT terms. First, we'll consider the statistics of certain sums of operators in the fixed point theory and argue that bulk memory is related to a failure of the central limit theorem. Second, we'll mock up modified bulk initial conditions as an RG transient and look for memory in the UV. In both cases, we'll find a sharp change of behavior at $\Delta = d/2$.

Consider, then, a CFT in $d$ dimensions, and focus on a patch of volume $L^d$, with a lattice cutoff at scale $\epsilon$. Choose an operator $O_i$ with dimension $\Delta_i$, and define the variable
\be
M_\epsilon = \sum_x O_i(x).
\ee
This sum contains one term per lattice point in the patch, for a total of $(L/\epsilon)^d$ summands, so as the lattice becomes small, the quantity $M_\epsilon$ involves a very large number of operators. This quantity is analogous to the Ising magnetization at level $u$ in the tree, with $p^{-u}\sim \epsilon/L$. Based on the second-eigenvalue condition, we expect the existence of memory to be connected with a transition as a function of $\Delta_i$. Specifically, we expect that for $\Delta_i > d/2$, small $\epsilon$ will make the distribution $P(M_\epsilon)$ Gaussian, while for $\Delta_i < d/2$ the distribution will remain non-Gaussian in the limit of small $\epsilon$. 

To confirm this, we will inspect some moments of the distribution for $M_\epsilon$. The two point function is
\be
\langle M_\epsilon^2\rangle = \sum_{x,y}\langle O_i(x)O_i(y)\rangle.
\ee
We can use translation invariance (ignoring a small correction due to finite volume) to do one of the sums,
\be
\langle M_\epsilon^2\rangle \approx \frac{L^d}{\epsilon^d}\sum_x\langle O_i(x)O_i(0)\rangle.
\ee
First, suppose $\Delta_i > d/2$. Then, the remaining sum is UV divergent, and the leading contribution is just the two point function at lattice scale. We normalize the operators so this is one, making the overall answer proportional to $(L/\epsilon)^d$. 

Next, suppose $\Delta_i < d/2$. Then the sum is dominated by the IR, so it can be approximated as an integral
$$
\int_0^L \frac{d^dx}{\epsilon^d} \frac{\epsilon^{2\Delta_i}}{|x|^{2\Delta_i}} \sim \frac{L^{d-2\Delta_i}}{\epsilon^{d-2\Delta_i}}.
$$
This means the two point function is proportional to $(L/\epsilon)^{2d - 2\Delta_i}$. To get a nicely normalized quantity in the continuum limit, we should divide $M_\epsilon$ not by the square root of the number of points, but by a fractional power, $M_\epsilon (\epsilon/L)^{d-\Delta_i}$.

To go further, consider the fourth moment, 
\be
\langle M_\epsilon^4\rangle = \sum_{x,y,z,w}\langle O_i(x)O_i(y)O_i(z)O_i(w)\rangle.
\ee
This is a four-point function, which depends on the entire operator spectrum of the theory. However, for $\Delta_i > d/2$, it is dominated by UV singularities when two pairs of the operators approach each other. There are three ways the operators can pair up, so, for large $L/\epsilon$
\be
\label{cumulant}
\langle M_\epsilon^4\rangle = 3\langle M_\epsilon^2\rangle\langle M_\epsilon^2\rangle.
\ee
This equation implies that the fourth order cumulant is zero, consistent with Gaussian statistics. Suppose, instead, that $\Delta_i < d/2$. Then there are no coincident-point divergences in the sum. It follows from scaling that result has to be proportional to $(L/\epsilon)^{4d-4\Delta_i}$. The coefficient depends on the OPE constants and spectrum of the theory, so, in general, the fourth-order cumulant will be nonzero.

Perhaps a more direct way to understand memory in a CFT is to study an RG transient, rather than focusing on the fixed point statistics. A related issue was considered in \cite{threefaces}, and we will use a similar construction. Specifically, we will mock up the bulk transient by adding an operator $O_i$ to the action, far from our patch of size $L^d$.\footnote{We are not smearing the operator over the $L^d$ patch as in \cite{threefaces}; we are inserting the operator at a definite location far away. Had we smeared the operator, we would have found an effect that becomes large in the UV for irrelevant $O_i$, unlike the bulk transient we are trying to model. We are grateful to Stephen Shenker for a discussion of this point.} Let us arrange the coefficient so that one point function of $O_i$ at the center of the patch, and renormalized at scale $L$, is order one. Within the patch, this means that the one point function of the operator $O_i(x)$ at the lattice scale is order $(\epsilon/L)^\Delta_i$, so the operator $M_\epsilon$ will have a one-point function of order $(\epsilon/L)^{\Delta_i -d}$. Comparing this to the variance of the distribution for $M_\epsilon$ in the unperturbed CFT, we find that the statistics of $M_\epsilon$ can detect the perturbation if $\Delta_i < d/2$, but not if $\Delta_i > d/2$.

\section{Super-horizon wave function in de Sitter space}\label{deSitterCal}

In this appendix, we review the calculation of the super-horizon wave function for a free massive field in the Bunch-Davies vacuum of de Sitter space, following \cite{maldacena} and \cite{Harlow:2011ke}. The wave function $\Psi$ depends on the spatial field $\phi(\mathbf{x})$, or its Fourier transform $\varphi_{\mathbf{k}}$, at a given conformal time $\eta_0$. It is given by a path integral over field configurations that satisfy a vacuum condition in the asymptotic past, and are equal to $\phi(\mathbf{x})$ at time $\eta_0$. As always in Gaussian theories, we can do this path integral by evaluating the action on the appropriate solution $\phi(\mathbf{x},\eta)$ of the equations of motion,
\be
\Psi[\eta_{0},\phi(\mathbf{x})]\propto e^{-iS_{cl}[\phi(\mathbf{x},\eta)]}.
\ee

As the in the main body of the paper, we'll work in flat slicing, with metric
\begin{equation}
ds^{2}=\ell_{dS}^{2}\frac{-d\eta^{2}+dx^{i}dx^{i}}{\eta^{2}} \hspace{20pt} i=1,\ldots,d,
\end{equation}
with $\ell_{dS}$ the de Sitter radius and $-\infty<\eta<0$.  With this metric, the action for a free massive scalar is 
\be
S = \frac{\ell_{dS}^{d+1}}{2}\int \frac{d^d\mathbf{x} d\eta}{\eta^d}\Big\{\frac{\eta^2}{\ell_{dS}^2} (\partial_\eta \phi)^2 - \frac{\eta^2}{\ell_{dS}^2}(\partial_i\phi)^2 - m^2\phi^2 \Big\}.
\ee
As before, we will introduce an explicit IR cutoff by making the identification $x^{i}\sim x^{i}+L$. This allows us to decompose the field into spatial Fourier components, $\phi(\eta,\mathbf{x})=\varphi_{\mathbf{k}}(\eta)e^{i\mathbf{k}\cdot\mathbf{x}}$ with quantized $\mathbf{k}=\frac{2\pi\mathbf{n}}{L}$. The equations of motion decouple into equations for each $\mathbf{k}$
\begin{equation}
\partial_\eta^2\varphi_{\mathbf{k}}-\frac{d-1}{\eta}\partial_\eta\varphi_{\mathbf{k}}+\left(\frac{m^{2}\ell_{dS}^{2}}{\eta^2}+k^{2}\right)\varphi_{\mathbf{k}}=0.\label{eq:de-sitter-massive-wave-equation}
\end{equation}
This differential equation is related to Bessel's equation and has the general solution
\begin{equation}
\varphi_{\mathbf{k}}(\eta)=\eta^{d/2}\left(A_{1}H_{\nu}^{(1)}(k\eta)+A_{2}H_{\nu}^{(2)}(k\eta)\right)
\end{equation}
\begin{equation}
\nu=\frac{1}{2}\sqrt{d^{2}-4m^{2}\ell_{dS}^{2}} = \frac{d}{2} - \Delta_-,
\end{equation}
where $H_{\nu}^{(1)}(k\eta)$ and $H_{\nu}^{(2)}(k\eta)$ are Hankel functions.

One linear combination of $A_1$ and $A_2$ is fixed by requiring that at time $\eta_0$, the solution should be equal to the Fourier transform of the argument of the wave function, $\varphi_{\mathbf{k}}$. Fixing the other linear combination amounts to making a choice of vacuum. We pick the Bunch-Davies vacuum \cite{davies}, also known as Hartle-Hawking \cite{Hartle:1983ai}, Eulidean or adiabatic. The prescription is to choose a solution that's purely positive frequency in the asymptotic past, $\eta\rightarrow-\infty$. This condition is made simple by the nice asymptotic properties of the Hankel function,
\begin{align}
\underset{|x|\to\infty}{\lim}&H_{\nu}^{(1)}(x)\sim e^{ix} \notag \\
\underset{|x|\to\infty}{\lim}&H_{\nu}^{(2)}(x)\sim e^{-ix}.
\end{align}
We recognize the latter as the positive frequency modes, so the correct solution is
\begin{equation}
\varphi_{\mathbf{k}}(\eta)=\varphi_{\mathbf{k}}\frac{\eta^{d/2}H_{\nu}^{(2)}(k\eta)}{\eta_{0}^{d/2}H_{\nu}^{(2)}(k\eta_{0})}.\label{eq:de-sitter-classical-solution}
\end{equation}

All that remains is to substitute this solution into the action, mode by mode. We can integrate by parts and use the fact that Eq.~\eqref{eq:de-sitter-classical-solution} satisfies the equations of motion to reduce the $\eta$ integral to a boundary term at $\eta_0$:\footnote{A priori, there is also an oscillatory contribution from $\eta\rightarrow -\infty$. We kill this piece in the usual way, by rotating the contour for $\eta$ slightly. The condition that the early-time mode is positive frequency ensures that this contribution is exponentially suppressed at early imaginary time.}
\be
S_{cl} = \frac{L^{d}\ell_{dS}^{d-1}}{2}\sum_{\mathbf{k}}\eta_{0}^{1-d}\varphi_{-\mathbf{k}}\partial_\eta\varphi_{\mathbf{k}}(\eta)\Big|_{\eta=\eta_{0}}.\label{eq:action-boundary-term}
\ee
We can evaluate the derivative using a Hankel function identity
\begin{equation}
\label{identity}
\partial_\eta\varphi_{\mathbf{k}}(\eta)\Big|_{\eta=\eta_{0}}=\left(\frac{d-2\nu}{2\eta_{0}}+k\frac{H_{\nu-1}^{(2)}(k\eta_{0})}{H_{\nu}^{(2)}(k\eta_{0})}\right)\varphi_{\mathbf{k}}.
\end{equation}
We are interested in the wave function for superhorizon modes, for which $k\eta_0$ is much less than one, and the Hankel functions can be expanded as
\begin{align}
&H_{\nu}^{(2)}(x)\approx A(\nu)x^{\nu}+B(\nu)x^{-\nu}\hspace{20pt} (x\ll 1)\label{eq:hankel-function-limit}\\
&A(\nu)\equiv\frac{1-i\cot\nu\pi}{2^{\nu}\Gamma\left[1+\nu\right]} \hspace{20pt} B\left(\nu\right)\equiv\frac{2^{\nu}i\Gamma[\nu]}{\pi} \notag.
\end{align}

At this point, it is useful to focus on the square of the wave function, $|\Psi|^{2}$. This allows us to discard real terms in the action, since they contribute only a phase to $e^{-iS_{cl}}$. Using the expansion above, the assumption $\nu > 0$ and Euler's reflection formula for the $\Gamma$ function, we find
\begin{equation}
Re(-iS_{cl})=-L^{d}\ell_{dS}^{d-1}\sum_{\mathbf{k}}\left(\frac{k^{2\nu}}{\eta_{0}^{d-2\nu}}\right)\varphi_{\mathbf{k}}\varphi_{-\mathbf{k}}\frac{\Gamma[1-\nu]}{2^{2\nu}\Gamma[\nu]}\sin\nu\pi,
\end{equation}
so that, finally,
\begin{align}
|\Psi|^{2}&\sim\exp\Big\{-2\sum_{\mathbf{k}}\beta(k,\eta_0)|\varphi_{\mathbf{k}}|^2\Big\}  \\
\beta(k,\eta)&=\sin\nu\pi\frac{\Gamma[1-\nu]}{2^{2\nu}\Gamma[\nu]}\left(\frac{L^{d}\ell_{dS}^{d-1}k^{2\nu}}{\eta^{d-2\nu}}\right) \hspace{20pt} (\nu >0) \notag.
\end{align}


\mciteSetMidEndSepPunct{}{\ifmciteBstWouldAddEndPunct.\else\fi}{\relax}
\bibliographystyle{utphys}
\bibliography{memory}{}

\ifx\mcitethebibliography\mciteundefinedmacro
\PackageError{utphys.bst}{mciteplus.sty has not been loaded}
{This bibstyle requires the use of the mciteplus package.}\fi
\providecommand{\href}[2]{#2}\begingroup\raggedright\begin{mcitethebibliography}{10}

\bibitem{Gibbons:1977mu}
G.~Gibbons and S.~Hawking, ``{Cosmological Event Horizons, Thermodynamics, and
  Particle Creation},''
\href{http://dx.doi.org/10.1103/PhysRevD.15.2738}{{\em Phys.Rev.} {\bfseries
  D15} (1977) 2738--2751}.

\bibitem{Hawking:1981fz}
S.~Hawking and I.~Moss, ``{Supercooled Phase Transitions in the Very Early
  Universe},''
\href{http://dx.doi.org/10.1016/0370-2693(82)90946-7}{{\em Phys.Lett.}
  {\bfseries B110} (1982) 35}.

\bibitem{Wald:1983ky}
R.~M. Wald, ``{Asymptotic behavior of homogeneous cosmological models in the
  presence of a positive cosmological constant},''
\href{http://dx.doi.org/10.1103/PhysRevD.28.2118}{{\em Phys.Rev.} {\bfseries
  D28} (1983) 2118--2120}.

\bibitem{Starobinsky:1982mr}
A.~A. Starobinsky, ``{Isotropization of arbitrary cosmological expansion given
  an effective cosmological constant},''
{\em JETP Lett.} {\bfseries 37} (1983) 66--69.

\bibitem{Marolf:2010nz}
D.~Marolf and I.~A. Morrison, ``{The IR stability of de Sitter QFT: results at
  all orders},'' \href{http://dx.doi.org/10.1103/PhysRevD.84.044040}{{\em
  Phys.Rev.} {\bfseries D84} (2011) 044040},
\href{http://arxiv.org/abs/1010.5327}{{\ttfamily arXiv:1010.5327 [gr-qc]}}.

\bibitem{Hollands:2010pr}
S.~Hollands, ``{Correlators, Feynman diagrams, and quantum no-hair in deSitter
  spacetime},''
\href{http://arxiv.org/abs/1010.5367}{{\ttfamily arXiv:1010.5367 [gr-qc]}}.

\bibitem{Guth:1980zm}
A.~H. Guth, ``{The Inflationary Universe: A Possible Solution to the Horizon
  and Flatness Problems},''
\href{http://dx.doi.org/10.1103/PhysRevD.23.347}{{\em Phys.Rev.} {\bfseries
  D23} (1981) 347--356}.

\bibitem{Linde:1981mu}
A.~D. Linde, ``{A New Inflationary Universe Scenario: A Possible Solution of
  the Horizon, Flatness, Homogeneity, Isotropy and Primordial Monopole
  Problems},''
\href{http://dx.doi.org/10.1016/0370-2693(82)91219-9}{{\em Phys.Lett.}
  {\bfseries B108} (1982) 389--393}.

\bibitem{Starobinsky:1980te}
A.~A. Starobinsky, ``{A New Type of Isotropic Cosmological Models Without
  Singularity},''
\href{http://dx.doi.org/10.1016/0370-2693(80)90670-X}{{\em Phys.Lett.}
  {\bfseries B91} (1980) 99--102}.

\bibitem{reviewofmeasure}
For recent reviews of progress in this direction, see below.

\bibitem{Freivogel:2011eg}
B.~Freivogel, ``{Making predictions in the multiverse},''
  \href{http://dx.doi.org/10.1088/0264-9381/28/20/204007}{{\em
  Class.Quant.Grav.} {\bfseries 28} (2011) 204007},
\href{http://arxiv.org/abs/1105.0244}{{\ttfamily arXiv:1105.0244 [hep-th]}}.

\bibitem{Salem:2011qz}
M.~P. Salem, ``{Bubble collisions and measures of the multiverse},''
  \href{http://dx.doi.org/10.1088/1475-7516/2012/01/021}{{\em JCAP} {\bfseries
  1201} (2012) 021},
\href{http://arxiv.org/abs/1108.0040}{{\ttfamily arXiv:1108.0040 [hep-th]}}.

\bibitem{broadcasting}
W.~Evans, C.~Kenyon, Y.~Peres, and L.~J. Schulman, ``Broadcasting on trees and
  the ising model,'' {\em Ann. Appl. Probab.} {\bfseries 10 (2)} (2000)
  410--433.

\bibitem{mooresnell}
T.~Moore and J.~L. Snell, ``A branching process showing a phase transition,''
  {\em Journal of Applied Probability} {\bfseries 16} no.~2, (1979) pp.
  252--260. \url{http://www.jstor.org/stable/3212894}.

\bibitem{kestenstigum}
H.~Kesten and B.~Stigum, ``Additional limit theorems for indecomposable
  multidimensional galton-watson processes.,'' {\em Ann. math. Statist.}
  {\bfseries 37} (1966) 1463--1481.

\bibitem{onthepurity}
P.~Bleher, J.~Ruiz, and V.~Zagrebnov, ``On the purity of the limiting gibbs
  state for the ising model on the bethe lattice,'' {\em Journal of Statistical
  Physics} {\bfseries 79} (1995) 473--482.
  \url{http://dx.doi.org/10.1007/BF02179399}. 10.1007/BF02179399.

\bibitem{Breitenlohner:1982jf}
P.~Breitenlohner and D.~Z. Freedman, ``{Stability in Gauged Extended
  Supergravity},''
\href{http://dx.doi.org/10.1016/0003-4916(82)90116-6}{{\em Annals Phys.}
  {\bfseries 144} (1982) 249}.

\bibitem{Klebanov:1999tb}
I.~R. Klebanov and E.~Witten, ``{AdS / CFT correspondence and symmetry
  breaking},'' \href{http://dx.doi.org/10.1016/S0550-3213(99)00387-9}{{\em
  Nucl.Phys.} {\bfseries B556} (1999) 89--114},
\href{http://arxiv.org/abs/hep-th/9905104}{{\ttfamily arXiv:hep-th/9905104
  [hep-th]}}.

\bibitem{symmetree}
D.~Harlow, S.~H. Shenker, D.~Stanford, and L.~Susskind, ``{Tree-like structure
  of eternal inflation: A solvable model},'' {\em Phys.Rev.} {\bfseries D85}
  (2012) 063516,
\href{http://arxiv.org/abs/1110.0496}{{\ttfamily arXiv:1110.0496 [hep-th]}}.

\bibitem{cc}
D.~Anninos and F.~Denef, ``{Cosmic Clustering},''
\href{http://arxiv.org/abs/1111.6061}{{\ttfamily arXiv:1111.6061 [hep-th]}}.

\bibitem{Magan:2013msa}
J.~M. Magán and A.~Sharma, ``{Memory as order-parameter on a Cayley Tree},''
\href{http://arxiv.org/abs/1304.7008}{{\ttfamily arXiv:1304.7008
  [cond-mat.stat-mech]}}.

\bibitem{mossel2001reconstruction}
E.~Mossel, ``Reconstruction on trees: beating the second eigenvalue,'' {\em The
  Annals of Applied Probability} {\bfseries 11} no.~1, (2001) 285--300.

\bibitem{dsreview}
For reviews of dS, see below.

\bibitem{Bousso:2002fq}
R.~Bousso, ``{Adventures in de Sitter space},''
\href{http://arxiv.org/abs/hep-th/0205177}{{\ttfamily arXiv:hep-th/0205177
  [hep-th]}}.

\bibitem{Anninos:2012qw}
D.~Anninos, ``{De Sitter Musings},''
  \href{http://dx.doi.org/10.1142/S0217751X1230013X}{{\em Int.J.Mod.Phys.}
  {\bfseries A27} (2012) 1230013},
\href{http://arxiv.org/abs/1205.3855}{{\ttfamily arXiv:1205.3855 [hep-th]}}.

\bibitem{Spradlin:2001pw}
M.~Spradlin, A.~Strominger, and A.~Volovich, ``{Les Houches lectures on de
  Sitter space},''
\href{http://arxiv.org/abs/hep-th/0110007}{{\ttfamily arXiv:hep-th/0110007
  [hep-th]}}.

\bibitem{Marolf:2010zp}
D.~Marolf and I.~A. Morrison, ``{The IR stability of de Sitter: Loop
  corrections to scalar propagators},''
  \href{http://dx.doi.org/10.1103/PhysRevD.82.105032}{{\em Phys.Rev.}
  {\bfseries D82} (2010) 105032},
\href{http://arxiv.org/abs/1006.0035}{{\ttfamily arXiv:1006.0035 [gr-qc]}}.

\bibitem{Witten:2001kn}
E.~Witten, ``{Quantum gravity in de Sitter space},''
\href{http://arxiv.org/abs/hep-th/0106109}{{\ttfamily arXiv:hep-th/0106109
  [hep-th]}}.

\bibitem{Strominger:2001pn}
A.~Strominger, ``{The dS / CFT correspondence},'' {\em JHEP} {\bfseries 0110}
  (2001) 034,
\href{http://arxiv.org/abs/hep-th/0106113}{{\ttfamily arXiv:hep-th/0106113
  [hep-th]}}.

\bibitem{Maldacena:2002vr}
J.~M. Maldacena, ``{Non-Gaussian features of primordial fluctuations in single
  field inflationary models},'' {\em JHEP} {\bfseries 0305} (2003) 013,
\href{http://arxiv.org/abs/astro-ph/0210603}{{\ttfamily arXiv:astro-ph/0210603
  [astro-ph]}}.

\bibitem{Freivogel:2006xu}
B.~Freivogel, Y.~Sekino, L.~Susskind, and C.-P. Yeh, ``{A Holographic framework
  for eternal inflation},''
  \href{http://dx.doi.org/10.1103/PhysRevD.74.086003}{{\em Phys.Rev.}
  {\bfseries D74} (2006) 086003},
\href{http://arxiv.org/abs/hep-th/0606204}{{\ttfamily arXiv:hep-th/0606204
  [hep-th]}}.

\bibitem{Heemskerk:2010hk}
I.~Heemskerk and J.~Polchinski, ``{Holographic and Wilsonian Renormalization
  Groups},'' \href{http://dx.doi.org/10.1007/JHEP06(2011)031}{{\em JHEP}
  {\bfseries 1106} (2011) 031},
\href{http://arxiv.org/abs/1010.1264}{{\ttfamily arXiv:1010.1264 [hep-th]}}.

\bibitem{Harlow:2011ke}
D.~Harlow and D.~Stanford, ``{Operator Dictionaries and Wave Functions in
  AdS/CFT and dS/CFT},''
\href{http://arxiv.org/abs/1104.2621}{{\ttfamily arXiv:1104.2621 [hep-th]}}.

\bibitem{mosselperes}
E.~Mossel and Y.~Peres, ``Information flow on trees.,'' {\em Ann. Appl.
  Probab.} {\bfseries 13 (3)} (2003) 817--844.

\bibitem{georgii}
H.~Georgii, {\em Gibbs Measures and Phase Transitions.}
\newblock de Gruyter, Berlin, 1988.

\bibitem{denef}
F.~Denef, ``{TASI lectures on complex structures},''
\href{http://arxiv.org/abs/1104.0254}{{\ttfamily arXiv:1104.0254 [hep-th]}}.

\bibitem{Benna:2011as}
M.~K. Benna, ``{De (Baby) Sitter Overlaps},''
\href{http://arxiv.org/abs/1111.4195}{{\ttfamily arXiv:1111.4195 [hep-th]}}.

\bibitem{zabrodin}
A.~Zabrodin, ``{Non-Archimedean strings and Bruhat-Tits trees},''
\href{http://dx.doi.org/10.1007/BF01238811}{{\em Commun.Math.Phys.} {\bfseries
  123} (1989) 463}.

\bibitem{threefaces}
D.~Harlow, S.~H. Shenker, D.~Stanford, and L.~Susskind, ``{The Three Faces of a
  Fixed Point},''
\href{http://arxiv.org/abs/1203.5802}{{\ttfamily arXiv:1203.5802 [hep-th]}}.

\bibitem{maldacena}
J.~M. Maldacena, ``{Non-Gaussian features of primordial fluctuations in single
  field inflationary models},'' {\em JHEP} {\bfseries 0305} (2003) 013,
\href{http://arxiv.org/abs/astro-ph/0210603}{{\ttfamily arXiv:astro-ph/0210603
  [astro-ph]}}.

\bibitem{davies}
N.~Birrel and P.~Davies, {\em Quantum fields in curved space.}
\newblock Cambridge Univ. Press, 1982.

\bibitem{Hartle:1983ai}
J.~Hartle and S.~Hawking, ``{Wave Function of the Universe},''
\href{http://dx.doi.org/10.1103/PhysRevD.28.2960}{{\em Phys.Rev.} {\bfseries
  D28} (1983) 2960--2975}.

\end{mcitethebibliography}\endgroup

\end{document}